\documentclass[pra, reprint,superscriptaddress]{revtex4-1}
\usepackage{amsmath}
\usepackage{amsfonts}
\usepackage{graphicx}
 \usepackage{physics}
 \usepackage[usenames]{color}

 \def\be{\begin{equation}} \def\ee{\end{equation}}
\def\bea{\begin{eqnarray}} \def\eea{\end{eqnarray}}

\begin{document}

	\date{\today}
	\title{Long Range Ordered Phase in a Quantum Heisenberg Chain with Interactions beyond Nearest Neighbor}
	
	\author{Zehan Li}
	\email{Equal contribution}
	\affiliation{Department of Physics and Astronomy, University of Pittsburgh, Pittsburgh, PA 15260, USA}
	
	\author{Sayan Choudhury}
	\email{Equal contribution; sayan.choudhury@pitt.edu}
	\affiliation{Department of Physics and Astronomy, University of Pittsburgh, Pittsburgh, PA 15260, USA}

	\author{W. Vincent Liu}
	\email{wvliu@pitt.edu}
	\affiliation{Department of Physics and Astronomy, University of Pittsburgh, Pittsburgh, PA 15260, USA}
	\affiliation{Wilczek Quantum Center, School of Physics and Astronomy and T. D. Lee Institute, Shanghai Jiao Tong University, Shanghai 200240, China}
        \affiliation{Shanghai Research Center for Quantum Sciences, Shanghai 201315, China}
        \affiliation{Shenzhen Institute for Quantum Science and Engineering and Department of Physics, Southern University of Science and Technology, Shenzhen 518055, China}
	
	\begin{abstract}
	Spin ensembles coupled to optical cavities provide a powerful platform for engineering synthetic quantum matter. Recently, we demonstrated that cavity mediated infinite range interactions can induce fast scrambling in a Heisenberg $XXZ$ spin chain (Phys. Rev. Research {\bf 2}, 043399 (2020)). In this work, we analyze the kaleidoscope of quantum phases that emerge in this system from the interplay of these interactions.  Employing both analytical spin-wave theory as well as numerical DMRG calculations, we find that there is a large parameter regime where the continuous $U(1)$ symmetry of this model is spontaneously broken and the ground state of the system exhibits $XY$ order. This kind of symmetry breaking and the consequent long range order is forbidden for short range interacting systems by the Mermin-Wagner theorem. Intriguingly, we find that the $XY$ order can be induced by even an infinitesimally weak infinite range interaction. Furthermore, we demonstrate that in the $U(1)$ symmetry broken phase, the half chain entanglement entropy violates the area law logarithmically. Finally, we discuss a proposal to verify our predictions in state-of-the-art quantum emulators.
	\end{abstract}

	\maketitle

\section{Introduction} 
In recent years, the rapid advancements in cavity QED technologies have propelled extensive investigations of emergent phenomena in quantum many-body systems with cavity induced long range interactions \cite{kockum2019ultrastrong,sheikhan2016cavity,landig2016quantum,blass2018quantum,sheikhan2019cavity,halati2019cavity,halati2017cavity,niederle2016ultracold,mivehvar2017disorder,mivehvar2019cavity}. These systems provide a promising platform for realizing quantum spin liquids \cite{chiocchetta2020cavity}, supersolids \cite{dogra2016phase,sundar2016lattice,chen2016quantum}, exotic superconductors \cite{schlawin2019bcavity,schlawin2019cavity,chakraborty2020non}, charge density waves \cite{chen2020extended}, quantum many-body scars \cite{chen2020persistent}, time crystals \cite{tucker2018shattered,kessler2019emergent,yang2021dynamical}, chaotic dynamical phases \cite{lerose2018chaotic,lerose2019impact}, and even topological states of matter \cite{sheikhan2016bcavity,wang2019cavity}. Moreover, cavity mediated interactions can be harnessed to explore many-body chaos \cite{bentsen2019integrable,bentsen2019treelike,marino2019cavity,alavirad2019scrambling,lewis2019unifying} and dynamical quantum phase transitions \cite{klinder2015dynamical,muniz2020exploring}.\\

In a recent paper, we have demonstrated that a one dimensional Ising spin chain coupled to a single mode cavity can exhibit fast scrambling; this highly chaotic dynamics originates from the interplay of short and long range interactions \cite{li2020fast}. Concurrently, other groups have also shown that competing short and long range interactions can induce fast scrambling \cite{belyansky2020minimal,yin2020bound}. In this context, it is worth noting that even though scrambling is an inherently non-equilibrium phenomenon, several fast scrambling many-body models host a rich array of quantum phases at equilibrium \cite{sachdev1993gapless,lunkin2018sachdev,song2017strongly,lantagne2018family,haldar2018higher,zhang2018topological}. This observation naturally leads to the following question: what are the ground state phases of this new class of cavity induced fast scramblers?  \\

\begin{figure}[h]
		\includegraphics[scale=0.33]{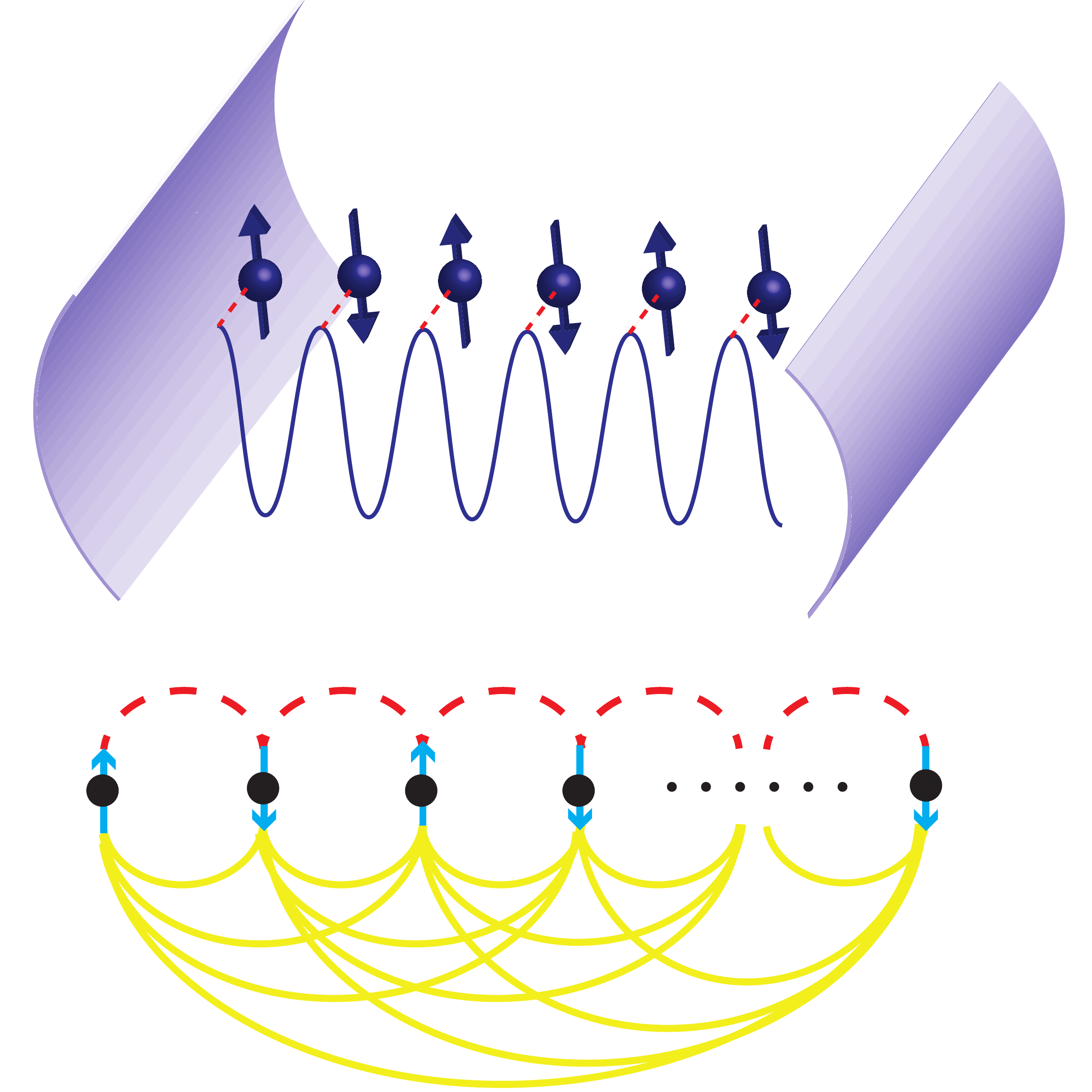}
		\caption{{\bf Schematic representation of the model:} The model in Eq.~(\ref{model}) is characterized by a nearest neighbor $XXZ$ coupling and an infinite range $XX$ coupling. This model describes a $XXZ$ spin chain coupled to a single mode cavity.}
		\label{fig_Model}
\end{figure}

In this paper, we address this question by investigating the quantum phases of an one-dimensional spin chain composed of two ingredients --- a nearest neighbor $XXZ$ interaction and an infinite range $XX$ interaction. A schematic representation of our model is shown in Fig.~\ref{fig_Model}. As shown in section V, this model describes a Heisenberg $XXZ$ spin chain coupled to a single mode cavity in the ``bad cavity" limit. By employing an analytical spin-wave analysis as well as numerical density matrix renormalization group (DMRG) computations, we demonstrate that this system exhibits three different phases: (a) a long-range ordered Ising ferromagnetic phase, (b) a quasi-long range ordered critical phase, and (c) a long-range ordered $U(1)$ symmetry breaking $XY$ phase. While the first two phases can be realized in the short range interacting Heisenberg model, the cavity induced interaction leads to the realization of the third phase. We demonstrate that these phases can be distinguished by their entanglement entropy; in particular, phases (b) and (c) violate the area law logarithmically and can be associated with an effective central charge. The effective central charge distinguishes phase (b) from phase (c).\\

This paper is organized as follows. In section II, we introduce our model and describe its ground states in two well known limits. In section III, we employ spin wave analysis to derive the phase diagram of this system. In section IV, we supplement the spin wave analysis with DMRG calculations on finite size chains. We discuss a potential experimental realization of our model in section V and conclude this paper in section VI with a summary of our findings.

\section{Model} 

We study a one dimensional spin chain with $N$ sites described by the Hamiltonian:
\begin{eqnarray}
	H  &=&- \frac{1}{4} \sum_{i=1}^{N-1}  \left(\sigma_i^z \sigma^z_{i+1} +\alpha(\sigma_i^x \sigma^x_{i+1}+\sigma_i^y \sigma^y_{i+1}) \right) \nonumber \\
    &+& \frac{J}{4 N}\sum_{i=1}^{N-1} \sum_{j>i} \left(\sigma_i^x \sigma^x_{j}+\sigma_i^y \sigma^y_{j} \right)  
	\label{model}
\end{eqnarray}
where $\sigma_i ^{\gamma}$ is the standard Pauli matrix at lattice site $i$. We have rescaled the infinite range interaction by $1/N$ to ensure extensivity of the total energy.\\

We note that this model is characterized by a $U(1) \times \mathbb{Z}_2$ symmetry. The $U(1)$ symmetry transformation operator is $M_U=\exp(-i \theta \sum_j \sigma_j^z)$, and it originates from the conservation of the total z-Magnetization. The breaking of this continuous symmetry implies that $\langle S_j^{+} \rangle \ne 0$ (where $S_j^{+} = (\sigma^x + i \sigma^y)/2$) and the system is in the $XY$ phase \cite{maghrebi2017continuous,peter2012anomalous}. On the other hand, the $\mathbb{Z}_2$ symmetry transformation operator is $\prod_j i \sigma_j^x$ (or $\prod_j i \sigma_j^y$), and it denotes a global rotation by $\pi$ about the $x$ (or $y$) axis. The chain is in the Ising ferromagnetic phase when the $\mathbb{Z}_2$ symmetry is broken and $\langle \sigma_j^z \rangle \ne 0$. \\

When $J \rightarrow 0$, the model reduces to the Heisenberg $XXZ$ model and it is the exactly solvable by the Bethe ansatz \cite{gaudin1971thermodynamics,destri1992new}. In this case there are two possible phases: the Ising ferromagnetic phase (when $\alpha < 1$) and a quasi-long range ordered critical phase, known as the Tomonaga-Luttinger Liquid (TLL) (when $\alpha \ge 1$) \cite{barmettler2010quantum}. We note that the Mermin-Wagner theorem forbids the existence of a truly long range ordered phase with only short range interactions \cite{mermin1966absence,giamarchi2003quantum}.\\

The ground state of this system can also be exactly determined in the $J \rightarrow \infty$ limit, when the model reduces to mean-field solvable Lipkin-Meshkov-Glick (LMG) model \cite{botet1982size,lipkin1965validity,ribeiro2007thermodynamical}. In this case, the ground state of the system is in the $XY$ phase \cite{vzunkovivc2016dynamical}. In the next section, we explore the phase diagram of this model when $J$ is finite. This is precisely the regime, where the model is non-integrable and its out-of-equilibrium dynamics is chaotic. \\

\section{Spin Wave Analysis} 
In this section, we employ spin-wave analysis to explore the phase diagram of the model. It is well known that the ground state spontaneously breaks the $\mathbb{Z}_2$ symmetry, when $\alpha \rightarrow 0$ and $J  \rightarrow 0$. In order to determine the phase boundary of this Ising ferromagnetic (FM) state, we define the vacuum state to be: \\
\begin{equation}\label{fmwf}
\vert \psi \rangle_{\rm FM} = \vert \uparrow  \uparrow  \uparrow  \uparrow \ldots  \uparrow  \uparrow  \uparrow  \uparrow \rangle,
 \end{equation}
 and apply the Holstein-Primakoff transformation to map the spin excitations to bosons: $S^-_j = \frac{1}{2}(\sigma_j^x - i \sigma_j^y)= \left(\sqrt{1-a_j^{\dagger}a_j}\right) a_j; S^+_j = \frac{1}{2}(\sigma_j^x + i \sigma_j^y) = a_j^{\dagger} \left(\sqrt{1-a_j^{\dagger}a_j} \right); S^z_j = (\frac{1}{2}-a_j^{\dagger}a_j)$ \cite{holstein1940field}. In the weak excitation regime, $\langle a^{\dagger}a \rangle \ll 1$, and the Hamiltonian describing these spin waves is given by: \\
\begin{eqnarray}
H_{FM} &=& \sum_i  \left( (a_i^{\dagger}a_i + a_{i+1}^{\dagger}a_{i+1}) - \alpha (a_i^{\dagger}a_{i+1}+a_{i+1}^{\dagger}a_{i}) \right) \nonumber\\
&+& \frac{J}{N}  \sum_i  \sum_{j>i} \left(a_i^{\dagger}a_{j}+a_{j}^{\dagger}a_{i}) \right)  
\end{eqnarray}
Assuming periodic boundary conditions, we can express the spin-wave Hamiltonian can be in the following form:
\begin{equation}
    H_{FM} = \sum_k \omega_k a_k^{\dagger}a_k,
\end{equation}
where 
\begin{equation}
\omega_k = 1-\alpha \cos(k) + \frac{J}{N} \sum_{r=1}^{N/2} \cos(\frac{2 \pi k}{N} r),
\end{equation}
where we have set the lattice constant to be $1$. \\

If ${\rm min} [\omega_k] > 0$, then the ground state of $H_{FM}$ is the vacuum state $\vert 0 \rangle$, such that
\begin{equation}
  a_k \vert 0 \rangle = 0 \,\, \forall \,\, k.
\end{equation}
In this case the ground state of our model is the $z$-polarized state described in Eq.~\ref{fmwf}. On the other hand, when ${\rm min} [\omega_k] < 0$, then the system is no longer in the weak excitation regime and the spin-wave approximation outlined above breaks down. Thus, the $z$-polarized state is not the correct choice for the quantum ground state in this regime, and the system exhibits instability towards $XY$ ordering. From these considerations, it is clear that the ground state is ferromagnetic when $\alpha = 1$ (for $J \ge 0$), and $\alpha=1+J/2$ (for $J \le 0$).\\

The Holstein-Primakoff transformation can also be employed to study the stability of the $U(1)$-symmetry breaking phase. In this case, we define the vacuum state to be spin polarized along the $+x$ direction:
\begin{equation}
\vert \psi \rangle_{\rm XY} = \vert \rightarrow  \rightarrow  \rightarrow  \rightarrow \ldots  \rightarrow  \rightarrow  \rightarrow  \rightarrow \rangle,
 \end{equation}

The Holstein-Primakoff mapping in this case is $S_i^x = (\frac{1}{2}-a_i^{\dagger}a_i)$; $S_i^y \approx a_i^{\dagger} + a_i$; $S_i^z \approx (a_i^{\dagger} - a_i)/i$. The Hamiltonian describing the spin-wave excitations in this case is:
\begin{equation}
H_{\rm sw} = \sum_{k=-N/2}^{N/2} \omega_k (a_k^{\dagger} a_k + a_{-k} a_{-k}^{\dagger}) + \mu_k (a_k^{\dagger} a_{-k}^{\dagger} + a_k a_{-k} );
\end{equation}
where,
\begin{eqnarray}
\omega_k&=& (\alpha - \frac{J}{2}) -\frac{1+\alpha}{2} \cos (\frac{2 \pi k}{N}) + \frac{J}{2N} \sum_{r=1}^{N/2} \cos (\frac{2 \pi k}{N} r) \nonumber\\
\\
\mu_k&=& \frac{1-\alpha}{2} \cos (\frac{2 \pi k}{N}) - \frac{J}{2N} \sum_{r=1}^{N/2} \cos (\frac{2 \pi k}{N} r) \
\end{eqnarray}
where $a_k = \frac{1}{\sqrt{N}} \sum_j \exp(i 2 \pi j k/N) a_j$. $H_{\rm sw}$ can be diagonalized by a Bogoliubov transformation \cite{takahashi1989modified}. In this case, the Bogoliubov quasiparticles are composed of both particles and holes and the ground state of the spin chain has spin excitations. The density of these excitations is given by:
\begin{eqnarray}
\langle a_i^{\dagger} a_i \rangle &=& \lim_{N \rightarrow \infty} \frac{1}{2 N} \sum_{k \ne 0} ([1- \mu_k^2/\omega_k^2]^{-1/2} -1) \nonumber\\
&=& \frac{1}{4 \pi} \int_{- \pi}^{\pi} dq \left([1- \mu(q)^2/\omega(q)^2]^{-1/2} -1 \right) \nonumber \\
&=& \frac{1}{4 \pi} \int_{- \pi}^{\pi} dq\,\,  {\mathcal I}(q)
\end{eqnarray}

\begin{figure*}
		\includegraphics[scale=0.45]{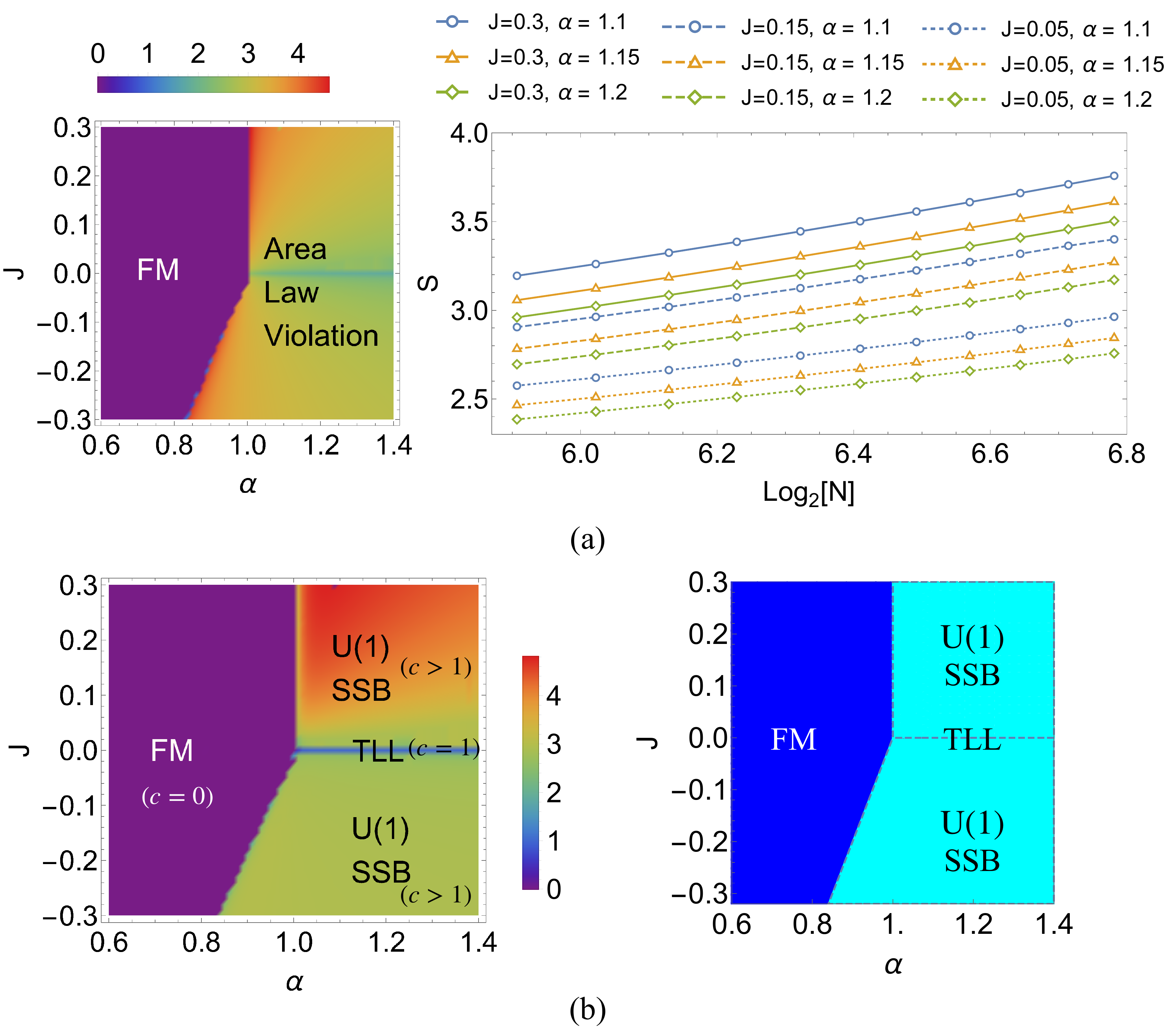}
		\caption{ (a) {\bf Ground State Entanglement Entropy:} The entanglement entropy, $S$ is $0$, when the spins are polarized along the z-direction and the correlations are ferromagnetic in nature. The entanglement entropy violates the area law logarithmically, when the correlations are $XY$-like. The left panel shows the density plot for the half chain entanglement entropy, defined in Eq.~\ref{EE} for a 100 site chain. The right panel shows the dependence of the entanglement entropy on the system size, when the correlations are $XY$-like. (b) {\bf Phase Diagram :} The left panel shows the phase diagram obtained from the effective central charge, $c$ (defined in Eq.~\ref{ccdef}). The density plot for $c$  reveals three phases: (1) A ferromagnetic phase characterized by $c=0$ (2) the critical TLL characterized by $c=1$ and (3) A true $U(1)$ spontaneous symmetry breaking (SSB) long range ordered phase characterized by $c>1$. The right panel shows the phase diagram obtained from spin wave analysis. The phase diagram obtained from both approaches match qualitatively. As mentioned in the main text, a purely mean-field analysis would misidentify the TLL phase as the $U(1)$ SSB phase.}
		\label{fig_EE}
\end{figure*}

By expanding the integrand around $q=0$, we find that $ {\mathcal I} (q) \propto 1/\sqrt{(J-\alpha q^2)(1-\alpha + (q^2-J)/2)}$, and  $ {\mathcal I} (q) \propto 1/\vert q \vert$, when $J=0$. This implies that in the absence of the infinite range interactions, $\langle a_i^{\dagger} a_i \rangle \sim \ln(N)$ and the long range order is destroyed in the thermodynamic limit; in this case, the system is in the quasi-long range ordered Tomonaga Luttinger Liquid (TLL) phase. On the other hand, $\langle a_i^{\dagger} a_i \rangle$ does not diverge and $U(1)$ symmetry breaking occurs $(S_j^{+} \propto e^{i \theta_0})$, when $J \ne 0$. This symmetry breaking and the suppression of the TLL phase originates from the mean-field nature of this model in the presence of infinite range interactions. Our results are summarized in Fig.~\ref{fig_EE}(b) (right panel). We note that while a mean-field analysis can correctly determine the phase boundary for the FM state, it would incorrectly identify the TLL phase as the $U(1)$-symmetry breaking phase in the $J=0$ regime. In the next section, we compliment our spin wave analysis results with numerical density matrix renormalization group calculation of the ground state phase diagram. \\

\section{Density Matrix renormalization Group Simulations}

The DMRG is a powerful tool to diagnose the equilibrium phases and out-of-equilibrium dynamics of one-dimensional and quasi-one-dimensional quantum systems \cite{white1992density,white1993density,white2004real}. We now proceed to  to determine the phase diagram of our model using the DMRG algorithm. In this method, we employ a matrix product state ansatz to represent the ground state \cite{DMRG, schollwock2011density}, and ensure that the algorithm converges globally with a truncation error less than $10^{-6}$. The short range part of the Hamiltonian (the $XXZ$ Heisenberg Model) has already been extensively studied with this method \cite{white1993density}. For the long range part, we represent $H_{LMG}$ as a sum of matrix product operators; this choice avoids systematic errors introduced by other schemes \cite{ren2020entanglement}. Our codes are mainly based on ${\it tensors.net}$ library~\cite{DMRGwebsite}.\\

The ground state entanglement entropy, provides a powerful tool to numerically diagnose the phases of long range interacting systems \cite{koffel2012entanglement,vodola2015long,eisert2010colloquium,cho2018realistic,kuwahara2020area,jiang2012identifying,isakov2011topological,zhang2012quasiparticle,vitagliano2010volume}. In particular, the $\mathbb{Z}_2$-symmetry broken ferromagnetic phase is characterized by an area law entanglement entropy, while ground states with $XY$-like order exhibit violation of the area law. We compute the entanglement entropy, $S$, defined as:
\begin{equation}\label{EE}
S= {\rm Tr} \rho_B \log(\rho_B),
\end{equation}
where $\rho_B$ is the reduced density matrix of the right  (left) half of the chain, and it is obtained by tracing over the degrees of freedom of the left (right) half of the chain. As shown in Fig.~\ref{fig_EE}(a) (left panel), $S=0$, when the spins are z-polarized and the spin chain is in the ferromagnetic phase. On the other hand, the entropy is finite, when the ground state is $XY$-like. \\

It is evident from Fig.~\ref{fig_EE}(a) (right panel) that in the $XY$-like phase, the entanglement entropy violates the area law logarithmically. Employing an analogy with critical systems \cite{holzhey1994geometric,calabrese2004entanglement}, we can define an effective central charge, $c$ using the following relation:\\
\begin{equation}\label{ccdef}
S= \frac{c}{6} \log(L) \\
\end{equation}

The central charge, $c$ is 0 for the Ising ferromagnetic phase and it is $1$ for the TLL phase. Furthermore, in the long range ordered $U(1)$ symmetry breaking $XY$ phase, $c>1$ \cite{gong2016kaleidoscope,maghrebi2017continuous,ren2020entanglement}. We note that the transition from the TLL phase to the $XY$ phase is a continuous Berezinskii-Kosterlitz-Thouless transition \cite{maghrebi2017continuous}. Thus, $c$ changes continuously when $J$ changes, and the area law is violated logarithmically in both phases. As shown in Fig.~\ref{fig_EE}(b) (left panel), we find that the cavity mediated long range interactions can lead to the spontaneous breaking of a continuous $U(1)$ symmetry for a large parameter regime. Furthermore, our results demonstrate that even an infinitesimally weak coupling between the short range interacting spin chain and the optical cavity is sufficient to induce long range $XY$ order in the spin chain, thereby providing a route to circumvent the Mermin-Wagner theorem.

\section{Proposed Experimental Realization}

As mentioned in the introduction, coupling a Heisenberg $XXZ$ spin chain to a single mode cavity provides a natural route to realize our model. The Heisenberg Hamiltonian can be engineered using Rydberg atoms \cite{whitlock2017simulating,nguyen2018towards,signoles2021glassy}, ultracold atomic gases \cite{duan2003controlling,jepsen2020spin,zhao2019engineered,pai2005superfluid}, and trapped ions \cite{davoudi2020towards}. In this section, we explicitly derive the effective spin Hamiltonian that arises when this scenario is realized.\\

The evolution of the density matrix of the system, $\hat{\rho}$ in the rotating frame of the atomic transition frequency can be described by the master equation:
\begin{equation}
         \frac{d\hat{\rho}}{dt}=-i[\hat{H}_{SL},\hat{\rho}]+\mathcal{L}_{c}[\hat{\rho}],
\end{equation}
where
\begin{equation}
         \hat{H}_{SL}= \Delta_c \hat{a}^+ \hat{a}+H_{XXZ}+g \sum_{i=1}^N (\hat{a}^+\hat{\sigma}_i^- + \hat{a}\hat{\sigma}_i^+).
\end{equation}

Here $\Delta_c$ is the detuning of the cavity mode frequency from the atomic transition frequency in the rotating frame, $g$ is the coupling between the atomic spins and the cavity field, $H_{XXZ}$ is the Heisenberg Hamiltonian described by:
\begin{equation}
         \hat{H}_{XXZ}= - \frac{1}{4} \sum_{i=1}^{N-1}  \left(J_z \sigma_i^z \sigma^z_{i+1} + J_{\rm xx} (\sigma_i^x \sigma^x_{i+1}+\sigma_i^y \sigma^y_{i+1}) \right) ,
         \label{xxz}
\end{equation}
and the photon loss from the cavity at a rate $\kappa$ is given by the Lindblad term:
\begin{equation}
         \mathcal{L}_c[\hat{\rho}]=\frac{\kappa}{2}(2\hat{a}\hat{\rho}\hat{a}^+ - \hat{a}^+\hat{a}\hat{\rho} - \hat{\rho}\hat{a}^+\hat{a}).
\end{equation}

By adiabatically eliminating the cavity mode in the bad cavity limit ($\kappa \gg g$), we obtain a master equation for the reduced density matrix $\hat{\rho}_s$ of the spin chain,

\begin{equation}
         \frac{d\hat{\rho_s}}{dt}=-i[\hat{H}_{{\rm eff}},\hat{\rho}_s]+\mathcal{L}_{\Gamma}[\hat{\rho}_s],
\end{equation}
where the effective Hamiltonian is given by:
     \begin{equation}
         \hat{H}_{{\rm eff}}=\frac{4 g^2 \Delta_c}{4\Delta_c^2 + \kappa^2} \sum_{i,j}\hat{\sigma}_i^+\hat{\sigma}_j^- + H_{XXZ},
     \end{equation}
and 
\begin{equation}
         \mathcal{L}_{\Gamma}[\hat{\rho}_s]= \frac{2 g^2 \kappa}{4\Delta_c^2 + \kappa^2}\sum_{i,j}(2\hat{\sigma}_i^-\hat{\rho}_s\hat{\sigma}_j^+ - \hat{\sigma}_i^+\hat{\sigma}_j^-\hat{\rho}_s - \hat{\rho}_s\hat{\sigma}_i^+\hat{\sigma}_j^-).
\end{equation}

 We conclude that the evolution of the spin chain is almost unitary when $\Delta_c \gg \kappa/2$; in this limit, the effective many-body model describing the system is given by Eq.~\ref{model}, with $\frac{J}{N} \approx \frac{4 g^2}{\Delta_c J_{\rm z}} $ and $\alpha = \frac{J_{\rm xx}}{J_z}$. \\
 
Interestingly, a highly tunable nearest-neighbor Heisenberg spin model has recently been realized with ultracold bosonic $^7$Li atoms loaded in an optical lattice \cite{jepsen2020spin}. In particular, near the Mott regime, the dynamics of this system is effectively described by $H_{XXZ}$ defined in Eq.~\ref{xxz}, where $J_{\rm xx} \sim 50$ Hz and $J_z/J_{\rm xx}$ can be tuned between $\sim -1.8$ and $\sim 1.6$. Furthermore, the infinite range interacting part of the Hamiltonian has also been emulated with cold atomic ensembles, where $g \sim 10$ kHz and $\Delta_c \sim 50$ MHz \cite{muniz2020exploring}. These results clearly demonstrate that the parameter regime of $\frac{J}{N}\sim 0.16$ appears well within the reach of on-going realistic experiments, thereby enabling the possibility of verifying our predictions in the near future.\\
 
\section{Summary and Outlook}
In this paper, we have examined the ground state phases of a Heisenberg spin chain with competing short and long range interactions. We have clearly demonstrated that cavity mediated infinite range interactions can lead to the spontaneous breaking of the continuous $U(1)$ symmetry and a consequent logarithmic violation of the area law. We have argued that the $U(1)$ symmetry breaking $XY$ phase can be identified by examining the effective central charge  of the ground state. Finally, we have outlined a proposal to realize our model in coupled cavity-quantum gas systems.\\

There are several future directions of this work. Firstly, it would be interesting to extend our study to spin-1 particles, and examine whether topological Haldane-like phases can arise in these systems. Furthermore, we can explore dynamical quantum phase transitions in these systems. Another promising direction would be to investigate the quantum phases and out-of-equilibrium dynamics of this model in various two dimensional geometries. Finally, we can also analyze the properties of this spin chain, when it is subjected to periodic driving. \\

\section*{Acknowledgments}
This work is supported by the AFOSR Grant No. FA9550-16-1-0006, the MURI-ARO Grant No. W911NF17-1-0323 through UC Santa Barbara, the Shanghai Municipal Science and Technology Major Project (Grant No. 2019SHZDZX01), and the University of Pittsburgh Center for Research Computing through the resources provided.

\bibliography{ref} 

\end{document}